\documentclass[11pt]{article} 
\usepackage{color}
\usepackage{url}
\usepackage{geometry} 
\usepackage{graphicx} 

\usepackage{float} 
\usepackage{wrapfig} 

\usepackage{lipsum} 

\usepackage{graphicx,times}
\usepackage{amssymb,amsmath}

\usepackage{supertabular}
 \usepackage{threeparttable}
 \usepackage{longtable}
\usepackage{color}
 \usepackage{multirow}
 \usepackage{url}
 \usepackage{txfonts}

\usepackage[a4paper=true,pagebackref=true]{hyperref}
\hypersetup{pdftitle = The title of my PDF, pdfauthor = My name, pdfsubject= The subject, pdfkeywords = keyword1 keyword2 keyword3} 
\hypersetup{colorlinks = true, linkcolor = green, anchorcolor = red, citecolor = blue, filecolor = red, pagecolor = red, urlcolor = red}

\begin{document}

\begin{center}
{\Large Gravitational wave research using pulsar timing arrays\\}
George Hobbs$^{1}$ \& Shi Dai$^{1}$ \\
$^1$ Australia Telescope National Facility, CSIRO, PO Box 76,  Epping. NSW 1710, Australia
\end{center}



A pulsar timing array (PTA) refers to a program of regular, high-precision timing observations of a widely distributed array of millisecond pulsars. Here we review the status of the three primary PTA projects and the joint International Pulsar Timing Array project. We discuss current results related to ultra-low-frequency gravitational wave searches and highlight opportunities for the near future.



%
\section{Introduction}           

Pulsar observations have been used for numerous astrophysical applications.  Not long after the discovery of pulsars, Counselman \& Shapiro (1968)~\cite{1968Sci...162..352C} described how observations of pulsars could be used to ``test general relativity, to study the solar corona, and to determine the earth's orbit and ephemeris time \ldots [and to determine] the average interstellar electron density".  Most studies to date have concentrated on analysing observations of specific pulsars. For instance, observations of one pulsar may provide excellent tests of general relativity, whereas another pulsar will be observed to probe the solar corona.  

During 1982 the first millisecond pulsar was discovered~\cite{1982Natur.300..615B}.  A few hundred millisecond pulsars are now known.  Their rotation is significantly more stable than the normal pulsars and their pulse arrival times can both be measured and also predicted with high accuracy. Foster \& Backer (1990)~\cite{1990ApJ...361..300F} showed how a comparison of timing observations from \emph{multiple} millisecond pulsars (a spatial array of pulsars) could be used to provide a time standard, to detect perturbations of the Earth's orbit and to search for gravitational waves (GWs).  They initiated an observing program (which they termed a ``pulsar timing array program") to observe three pulsars using the National Radio Astronomy Observatory 43\,m telescope.  During 2004, a much larger pulsar timing array (PTA) project began with the Parkes 64\,m telescope and is known as the Parkes Pulsar Timing Array (PPTA).  This project is ongoing (an overview and the first data release was described by Manchester et al. (2013)~\cite{2013PASA...30...17M}) and now the team undertakes regular observations of 25 pulsars in three observing bands.  The North American Nanohertz Observatory for Gravitational Waves (NANOGrav) in North America~\cite{2013CQGra..30v4008M} was founded in 2007 and uses the Arecibo and Green Bank telescopes to observe 36 pulsars.   Observations are also carried out for 42 pulsars with the Sardinian, Effelsberg, Nancay, Westerbork and Jodrell Bank telescopes by the European Pulsar Timing Array (EPTA) project team~\cite{2013CQGra..30v4009K, 2016MNRAS.458.3341D}, which was also founded in 2007.  The three project teams combine their expertise and data sets as part of the International Pulsar Timing Array (IPTA)~\cite{2016MNRAS.458.1267V,2010CQGra..27h4013H}.

In this review article, we will concentrate on one aspect of pulsar timing array research: searching for GW signals. The first observational evidence for GWs came from observations of a binary pulsar system (PSR B1913+16).  That system was discovered by Hulse \& Taylor (1975)~\cite{1975ApJ...195L..51H} and was shown to be losing energy at exactly the rate predicted by the theory of general relativity for GW emission.  The first direct detection of GWs was recently made by the LIGO/Virgo collaboration.  During 2015 they detected two bursts of GW emission coming from the coalescence of stellar mass black holes~\cite{2016PhRvL.116f1102A}.  This exciting result has opened the field of observational GW astronomy, and pulsar timing array projects provide a complementary view of the gravitational wave sky.  Whereas the LIGO/Virgo detectors allow us to detect high-frequency GWs from stellar mass systems, the pulsar observations will allow us to detect ultra-low-frequency GWs from supermassive binary black holes.  In contrast to the Hulse \& Taylor (1975) work that provided evidence for GWs, the PTAs will enable a direct detection of GWs.  For completeness we note that space-based detectors (such as the Laser Interferometer Space Antenna; LISA; Amaro-Seoane et al. 2017~\cite{2017arXiv170200786A}) will be sensitive to GWs in a frequency range between the PTA experiments and LIGO.

In this paper we describe how GWs affect pulsar observations (\S~2).  We give a brief summary of current data sets (\S~3) and the techniques being applied to hunt for the signals (\S~4).  In \S~5 we summarise the results obtained to date and in \S~6 we highlight some of the current limitations of the data sets and techniques. In \S~7 we consider the future of pulsar timing arrays and conclude in \S~8.

\section{How GWs affect pulsar observations?}

Various authors~\cite{1970Natur.227..157K, 1975GReGr...6..439E} determined how GWs affect an electromagnetic signal propagating from an emitting object to a detector. Such calculations were applied to pulse arrival times from pulsars by Sazhin (1978)~\cite{1978SvA....22...36S} and  Detweiler (1979)~\cite{1979ApJ...234.1100D}.  A GW induces a fluctuation in the observed pulse frequency, $\delta \nu/\nu$, of:

\begin{equation}\label{eqn:GWdoppler}
\frac{\delta \nu}{\nu} = -H^{ij}\left[h_{ij}(t_e,x_e^i)-h_{ij}(t_e - D/c,x_p^i)\right]
\end{equation}
where $H^{ij}$ is a geometrical term that depends on the position of the GW source, the Earth and the pulsar (at distance $D$ from the Earth).  The GW strain, $h_{ij}(t,x)$, is evaluated at the Earth (at time $t_e$ and position $x_e$) and at the pulsar (at the time the GW signal passed the pulsar, $t_p$, and at position $x_p$).  The shift of the pulse frequency is not directly measured.  Instead pulse times-of-arrival (ToAs) are determined.  These ToAs are then compared with predictions for the arrival times based on a pulsar timing model.  The differences between the predictions and the measurements are known as the pulsar ``timing residuals".  A GW signal will induce timing residuals at time $t$ from the initial observation of
\begin{equation}\label{eqn:GWres}
R(t) = -\int_0^t \frac{\delta \nu}{\nu} dt.
\end{equation}

The theory of general relativity predicts two polarisation states, $A_+$ and $A_\times$, for GWs (see Lee, Jenet \& Price 2008~\cite{2008ApJ...685.1304L} for non-GR predictions).  We can therefore re-write the Earth term as (note that the pulsar term is the same, but with an extra phase):
\begin{equation}
R_e(t) = \int_0^t \frac{P_+A_+(t) + P_\times A_\times(t)}{2(1-\gamma)} dt
\end{equation}
in which $P_+$ and $P_\times$ are geometrical terms and $\gamma$ is the GW-Earth-pulsar angle.  For a non-evolving, continuous wave source (i.e., from a non-evolving, supermassive, binary black hole system), the $A_{+,\times}$ will oscillate with an angular frequency of the GWs being $\omega_g$.  For a supermassive, circular, binary, black hole system the GWs will be emitted at twice the orbital frequency.  Eccentric binaries radiate GWs over a spectrum of harmonics of the orbital frequency. 

An estimation of the amplitude of the induced timing residuals caused by a binary system  can be determined from~\cite{2009arXiv0909.1058J}:
\begin{equation}
\Delta t \sim 10{\rm ns} \left(\frac{1{\rm Gpc}}{d}\right)\left(\frac{M}{10^9 {\rm M}_\odot}\right)^{5/3}\left(\frac{10^{-7}{\rm Hz}}{f}\right)^{1/3}
\end{equation}
where $d$ is the luminosity distance to the system which has a total mass of $M/(1+z)$ (where $z$ is the redshift) and the GW frequency is $f$.  More details are provided in Rosado et al. (2016)~\cite{2016PhRvL.116j1102R} who considered the detectability of binary systems at high redshift. They showed that very high mass ($> 10^{10}$\,M$_\odot$) binary systems could be detected by current PTAs at arbitrarily high redshifts.

\begin{figure}
\begin{center}
\includegraphics[width=8cm]{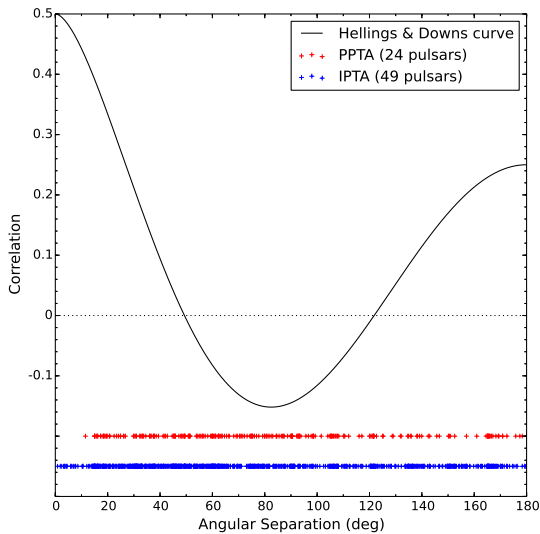}
\caption{The Hellings-and-Downs curve. The red dots indicate the angular separations of the pulsars in the PPTA project. The blue dots indicate the angular separations for the pulsars in the IPTA project.  Note that the IPTA provides coverage at all angular scales. }\label{fg:hdCurve}
\end{center}
\end{figure}

Of course, our universe will contain a large number of supermassive, binary black hole systems.  To determine the total GW signal from these systems we therefore need to sum Equation~\ref{eqn:GWdoppler} over all the sources.  This results in a background of GW signals. For an isotropic, stochastic, unpolarised background signal, Hellings \& Downs (1983)~\cite{1983ApJ...265L..39H} showed that the timing residuals for each pulsar pair will be correlated as:
\begin{equation}
c(\theta) = \frac{3}{2}x\ln x - \frac{x}{4} + \frac{1}{2} + \frac{1}{2}\delta(x)
\end{equation}
where $x = [1-\cos \theta]/2$ for an angle $\theta$ on the sky between two pulsars and $\delta(x)$ is the Dirac delta function\footnote{As described by Zhu (2015)~\cite{XingjiangPhd} this function is a factor of 3/2 larger than the original Hellings \& Downs (1983) result because of different scaling factors. Many publications parameterise the curve in different ways; see Jenet \& Romano (2015)~\cite{2015AmJPh..83..635J} for  a pedagogical discussion of the curve.}.  This analytic expression is plotted in Figure~\ref{fg:hdCurve} and is commonly referred to as the Hellings-and-Downs curve\footnote{Ravi et al. (2012)~\cite{2012ApJ...761...84R} showed how the curve would change for a relatively small number of sources and Lee, Jenet \& Price (2008)~\cite{2008ApJ...685.1304L} determined the expected correlations for general theories of gravity.}.  When searching for a background of GWs, the PTA teams therefore determine how correlated the timing residuals are for each pulsar pair.  A convincing detection of the GW background will be made if those correlations are shown to follow the Hellings-and-Downs curve.

Typically pulsars are observed every few weeks and the longest data spans are now a few decades (millisecond pulsars were discovered in 1982).  This implies that PTA data sets are sensitive to GWs with wavelengths from weeks to years.  These correspond to ultra-low-frequency ($10^{-9}$--$10^{-8}$\,Hz) GWs\footnote{Kopeikin (1997) showed that binary pulsars could potentially be used to detect even lower frequency GWs ($10^{-11}$--$10^{-9}$\,Hz) and  Dolch et al. (2016)~\cite{2016JPhCS.716a2014D} showed that specific observing campaigns can be carried out to search for GWs in the $10^{-6}$--$10^{-3}$\,Hz regimes. However, almost all of the work carried out so far has been in the ultra-low-frequency regime.}.

\section{Observations and timing data sets}

The three PTAs carry out regular timing observations of their sample of millisecond pulsars. Details of the observing systems have been presented in the various data release papers (see Desvignes et al. 2016~\cite{2016MNRAS.458.3341D}, Arzoumanian et al. 2015~\cite{2015ApJ...813...65T} and Manchester et al. 2013).  In brief, the data from a given telescope is generally folded online using the known timing model  for the pulsar being observed.  The resulting data files are processed to remove radio-frequency-interference and to apply various calibration procedures (such as removing instrumental delays and calibrating the polarisation and flux density of the signal).  Pulse ToAs are determined for each observation by cross-correlating the observed pulse profile with a template providing a high S/N representation of the expected pulse shape.

One of the primary noise sources that affect searches for GWs are variations in electron densities in the interstellar medium (see, for example, Keith et al. 2013~\cite{2013MNRAS.429.2161K} and Lee et al. 2014~\cite{2014MNRAS.441.2831L}).  Such changes can be monitored and (to some extent) removed or modelled by observing the pulsars over a wide range of frequencies.  The PPTA currently uses a dual band receiver providing simultaneous observations in the 10\,cm (3\,GHz) and 40\,cm (700\,MHz) observing bands along with a 20\,cm receiver (1400\,MHz).  The EPTA uses their large number of telescopes to obtain observations of each pulsar at frequencies between $\sim$300\,MHz with the Westerbork Synthesis Radio telescope and $\sim$2.6\,GHz with the Effelsberg radio telescope. Data in the 20\,cm observing band from five of the European telescopes are also combined as part of the Large European Array for Pulsars (LEAP) to form a tied-array telescope with an effective aperture equivalent to a 195\,m diameter telescope~\cite{2016MNRAS.456.2196B}. NANOGrav carries out observations between $\sim$300\,MHz and 2.4\,GHz.  

Some of the pulsars in the Southern hemisphere can only be observed by the PPTA.  The Northern hemisphere pulsars are generally observed by a large number of telescopes in both Europe and North America.  A few pulsars are observed by all three PTAs.  This has led to some observing campaigns in which a large number of IPTA telescopes observe the same source.  For instance, PSR~J1713$+$0747 was observed non-stop for 24\,hours (Dolch et al. 2016). A list of publically-accessible timing array data sets is given in Table~\ref{sec:data}.  It is expected that new data sets will continue to be released from the IPTA and the three PTA projects on a regular basis. 

\begin{table}
\caption{Publically available PTA data sets}\label{sec:data}
\begin{tabular}{lll}
\hline
PTA & Access & Reference \\ 
\hline
IPTA & \url{http://www.ipta4gw.org/?page_id=519} & Verbiest et al. (2016)~\cite{2016MNRAS.458.1267V}\\
NANOGrav & \url{https://data.nanograv.org} & Arzoumanian et al. (2015)~\cite{2015ApJ...813...65T}\\
EPTA & \url{http://www.epta.eu.org/aom.html} & Desvignes et al. (2016)~\cite{2016MNRAS.458.3341D}\\
PPTA & \url{http://doi.org/10.4225/08/561EFD72D0409} & Reardon et al. (2016)~\cite{2016MNRAS.455.1751R}\\
\hline
\end{tabular}
\end{table}

Even though they are not official members of the IPTA, other telescopes are used to observe millisecond pulsars and are likely to contribute to PTA research.  In China, the Nanshan, Yunnan, Shanghai and Jiamusi telescopes observe pulsars at a wide range of observing frequencies. The GMRT in India, LOFAR in Europe and the MWA in Australia observe pulsars at low frequencies (10 to 240\,MHz and 80 to 300\,MHz respectively).   Papers related to PTA research have also been published using observations from Kalyazin in Russia (e.g., Ilyasov et al. 2004\cite{2004IAUS..218..433I}).

\section{Techniques}
 
GWs will induce timing residuals.  The form of those timing residuals will depend on the nature of the GWs (single, non-evolving sources will induce sinusoidal residuals, a background will induce timing residuals that have a power-law spectrum). The statistical challenge is therefore to first search for statistically significant residuals and then to prove that they arise because of GWs.  If no GWs are detected then upper bounds on the GW amplitude can be determined.
 
 All existing techniques are based on the ``pulsar timing method".  Timing software packages (such as \textsc{tempo}, \textsc{tempo2} or \textsc{pint}\footnote{\textsc{tempo}, \textsc{tempo2} and \textsc{pint} are accessible from \url{http://tempo.sourceforge.net}, \url{https://bitbucket.org/psrsoft/tempo2} and \url{https://github.com/nanograv/PINT} respectively.}) are used to compare the measured pulse ToAs with predictions for those ToAs based on a model for the astrometric, pulse and, interstellar medium and orbital parameters of each pulsar.   Both ``frequentist" and ``Bayesian" methods exist for searching GWs.  We summarise these methods in Table~\ref{tb:methods}.  The various algorithms described in the table can be split into routines for specific types of GW signals (such as the \textsc{findCW} and \textsc{detectGWB} plugins to \textsc{tempo2}) or more general codes that can search for various GW types.  In all these cases, the user provides a set of high precision pulsar observations and defines the type of GW signal to be searched for along with information on what other noise processes are likely to be present in the data.    For instance, the measured timing residuals are not only induced by GWs as   pulse ToAs are affected by many phenomena.  Along the line-of-sight to the pulsar the interstellar medium and the Solar wind can contribute significant delays to the measured ToAs.  The measurement of a ToA is also sensitive to instrumental errors and incomplete polarisation calibration.  A GW background produces low-frequency timing residuals, but so do errors in terrestrial time standards, intrinsic pulsar instabilities and much more.  A detection of a GW signal therefore requires the confirmation of the expected spatial angular signature (the Hellings-Downs curve for a background or a quadrupolar spatial signature for a single GW source).  When searching for GWs it is therefore necessary either to first determine and then remove the non-GW noise processes or to search for the GWs whilst simultaneously modelling these other phenomena that affect the pulse arrival times.
  
Bayesian algorithms can be significantly slower than frequentist algorithms and GW searches using large data sets often require high-performance-computing facilities\footnote{A single step in the Bayesian algorithms may be just as fast as the computation of a frequentist statistic, however, the Bayesian methods sample a large parameter space of signals and noise processes.}. Frequentist-based methods often require that each noise process is dealt with in turn and, without care, this can lead to correlations between the different processes being unaccounted for in the final results. All the available algorithms should be used with care and tested on simulated data sets that have similar properties to the real data (i.e., different noise properties, irregular sampling, different data spans for different pulsars, etc.).  The IPTA produced a set of simulated data sets\footnote{Available from \url{http://www.ipta4gw.org/?page_id=89}.} with the primary goal of testing and comparing different GW detection algorithms.
  
 \begin{table*}
 \caption{Software routines that are publically available for GW searches}\label{tb:methods}
\begin{footnotesize} \begin{tabular}{p{1.0cm}p{8cm}p{4cm}p{3cm}}
\hline
Software & Access & Application & Reference \\
\hline
NX01 & \url{http://stevertaylor.github.io/NX01/} & Isotropic and anisotropic GW background, bursts with memory and individual source searches. &  Taylor (2017)\cite{nx01}\\
PAL2 & \url{https://github.com/jellis18/PAL2} & Isotropic and anisotropic GW background, bursts with memory, burst events and continuous waves & Ellis \& van Haasteren (2017)\cite{justin_ellis_2017_251456} \\
Piccard & \url{https://github.com/vhaasteren/piccard} & Isotropic and anisotropic GW background, bursts with memory, burst events and continuous waves & van Haasteren (2016)\cite{2016ascl.soft10001V} \\
Tempo2 & \url{https://bitbucket.org/psrsoft/tempo2} & Generic search for GW signal with arbitrary waveform & Madison et al. (2016)\cite{2016MNRAS.455.3662M} \\
									      & & Global least squares fitting for memory event & Wang et al. (2015)\cite{2015MNRAS.446.1657W} \\
Tempo2 plugins & Access with tempo2 distribution & findCW: search for continuous wave sources & Zhu et al. (2014)\cite{2014MNRAS.444.3709Z} \\
                           &                                                    & detectGWB: fast, but simple method for GWB detection & Tiburzi et al. (2016)\cite{2016MNRAS.455.4339T} \\
Temponest & \url{https://github.com/LindleyLentati/TempoNest} & Bayesian analysis tools including GWB searches & Lentati et al. (2014)\cite{2014MNRAS.437.3004L} \\ 
libstempo & \url{https://github.com/vallis/mc3pta/tree/master/stempo} & Routines for interfacing many of the software packages above with \textsc{tempo2}, \textsc{tempo} and \textsc{pint} & Developed primarily by Vallisneri, M. \\
\hline
 \end{tabular}\end{footnotesize}
 \end{table*}
 
\section{Results to date}

\begin{figure}
\begin{center}
\includegraphics[width=8cm]{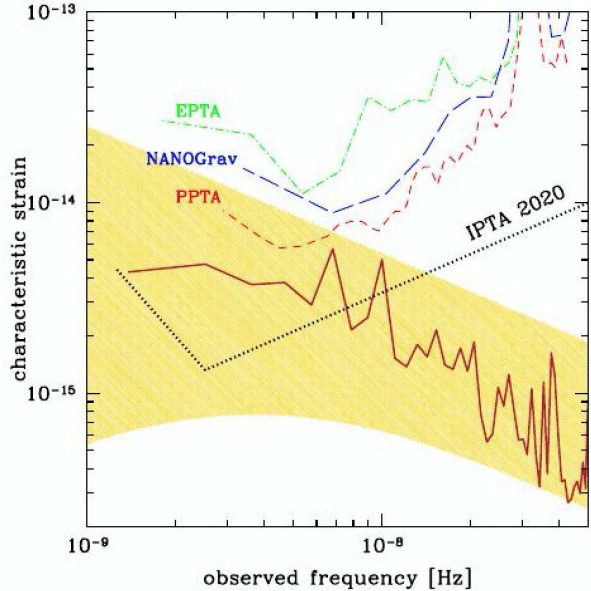}
\caption{Current upper bounds (dotted lines) and the theoretical expectation (shaded region and orange, solid curve) for the GW signal.  A possible bound that could be reached by the IPTA by 2020 is shown as the dotted line. Figure courtesy of A. Sesana.}\label{fg:prediction}
\end{center}
\end{figure}

Ultra-low-frequency GWs have not yet been detected.  Work has therefore been split between 1) predicting the expected signal and time to detection, 2) making more-and-more sensitive searches for the GWs and 3) understanding the astrophysical implications of our non-detections.  A summary is given in Figure~\ref{fg:prediction} where we show the current upper-bounds from the three PTAs as dotted lines.  A theoretical prediction (from Sesana et al. 2016) for the likely GW background signal from coalescing supermassive, binary black holes is shown in the shaded region. One possible realisation of such a background (made up from numerous individual black hole binaries) is shown as the jagged, solid line.  We note that the current PTAs are starting to constrain some models of the GW signal.  However, it is likely that we will require the sensitivity of the entire IPTA project to finally detect the GWs.  A bound that potentially could be reached by the IPTA within a decade or so is shown as the dotted, black line.  A more in-depth analysis of the likely time to detection has recently been published by Kelley et al. (2017).

In the following subsections we describe the results to date for different types of GW searches.  We note (and see e.g., Rosado et al. 2015~\cite{2015MNRAS.451.2417R}) that it is not yet clear whether we will first detect an individual supermassive binary black hole system or a background made up from a large number of GW sources.

\subsection{Individual supermassive binary black holes}

Lommen \& Backer (2001)~\cite{2001ApJ...562..297L} unsuccessfully searched for GW emission from Sagittarius A$^*$ that had been proposed to be a binary system.  They also searched for continuous GW emission from other nearby galaxies.  Sudou et al. (2003)~\cite{2003Sci...300.1263S} presented evidence for a supermassive black hole binary system in the radio galaxy 3C66B.  It was shown by Jenet et al. (2004)~\cite{2004ApJ...606..799J} that such a system would be producing detectable GWs and therefore ruled out the Sudou et al., models with high confidence.  Zhu et al. (2014)~\cite{2014MNRAS.444.3709Z}, Arzoumanian et al. (2014)~\cite{2014ApJ...794..141A} and Babak et al. (2016)~\cite{2016MNRAS.455.1665B} have presented recent bounds on individual GW sources over the entire sky or in particular sky directions with a representative value being that the GW strain amplitude:
\begin{equation}
h_0 < 2 \times 10^{-14} 
\end{equation}
at a frequency of 10\,nHz. The exact bound depends upon the sky direction and GW frequency.  Such bounds constrain the local merger rate density of supermassive binary black holes, but the current limits are higher than the theoretical expectations (e.g., Sesana 2013~\cite{2013CQGra..30x4009S} and Ravi et al. 2012~\cite{2012ApJ...761...84R}) for such GWs.

\subsection{Burst events}

Even though the first GW detection was a burst signal in the audio band (Abbott et al. 2016), the expectations for burst GW signals are not well developed in the pulsar timing band.  To date, the PTA teams have considered burst emission from the formation of supermassive black holes, highly eccentric black hole binaries, close encounters of massive objects, cosmic string cusps and memory events.  

Seto (2009)~\cite{2009MNRAS.400L..38S} showed how permanent distortions can occur in spacetime during mergers of supermassive binary black holes.  Such a ``memory event" will lead to a step change in the pulse frequency of all pulsars.  Cordes \& Jenet (2012)~\cite{2012ApJ...752...54C} gave an order-of-magnitude estimation of the signal-strength of:
\begin{equation}
h^{\rm mem} \sim 5 \times 10^{-16} \left(\frac{\mu}{10^8 {\rm M}_\odot}\right) \left(\frac{1 {\rm Gpc}}{D}\right)
\end{equation}
where $\mu$ is the reduced mass of the system and $D$ is the distance.  Searches for bursts with memory have been carried out by Arzoumanian et al. (2015) and Wang et al. (2015).  No detection was made and the current conclusion is that GW memory events are unlikely to be detected in the near future.  

\subsection{Background of GWs}

Most PTA research has been related to detecting or bounding the GW background.  It is common to define the background spectrum as:
\begin{equation}
h_c(f) = A\left(\frac{f}{f_{\rm 1yr}} \right)^\alpha
\end{equation}
where $f_{\rm 1yr} = 1/(1{\rm yr})$. The spectral exponent is thought to be close to $\alpha = -2/3$ for a background caused by coalescing black holes, $-1$ for cosmic strings and $-7/6$ for relic GWs (see Jenet et al. 2006~\cite{2006ApJ...653.1571J} and references therein).   

 All three PTAs (and the IPTA) are now placing bounds around $A < 10^{-15}$ with 95\% confidence (e.g., Shannon et al. 2015~\cite{2015Sci...349.1522S}, Lentati et al. 2015~\cite{2015MNRAS.453.2576L},   Arzoumanian et al. 2016~\cite{2016ApJ...821...13A}, Verbiest et al. 2016~\cite{2016MNRAS.458.1267V}).    The amplitude of the astrophysical GW background predicted by different models is based on various physical assumptions (see, e.g.,  Sesana 2013 for an overview). Firstly, models for the binary supermassive black hole (SMBH) population rely on measurements of the galaxy merger rate. Secondly, it has often been assumed that all galaxy mergers form binary SMBHs that coalesce well before a subsequent galaxy merger. Thirdly, the binary orbital decay has been assumed to be driven only by losses of energy to GWs when radiating in the pulsar timing frequency band. 

 The current GW background limits are inconsistent with some of the early theoretical models for the expected GW background signal, which suggests that at least one of the physical assumptions underlying such GWB models is likely to be incorrect. For instance, the galaxy-merger timescales might be longer than we currently expect. This would result in a lower merger rate and, hence, fewer binary SMBHs. It is also possible that not all large galaxies host SMBHs.  SMBH binaries may not efficiently reach the GW emitting stage in our assumed time scale (they ``stall"; e.g., Simon \& Burke-Spolaor 2016~\cite{2016ApJ...826...11S}) and other mechanisms in addition to GW emission drive binaries to coalescence, such as the coupling of binary SMBHs to their environments (e.g., Kocsis \& Sesana 2011~\cite{2011MNRAS.411.1467K}) or orbital eccentricity~\cite{2014MNRAS.442...56R, 2015PhRvD..92f3010H}). Binary SMBHs could lose energy and momentum because of three-body scattering of stars and viscous friction against circumbinary gaseous disk. Such coupling mechanisms also bend the GW background spectrum at low frequencies (i.e., large orbital separations).  Arzoumanian et al. (2016) checked the consistency of their limit with previously reported scaling relations between SMBH mass and galactic bulge mass, using fiducial estimates for galaxy merger rates and the stellar mass function. Under the assumption of circular GW-driven binaries, they found that the scaling relations of Kormendy \& Ho (2013)~\cite{2013ARA&A..51..511K} and McConnell \& Ma (2013)~\cite{2013ApJ...764..184M} to be inconsistent with their data at the 95\% and 90\% level respectively. They also placed constrains on the strength of environmental coupling effects via parameterisation of the GW background spectrum that allows for a turn-over frequency (see also Sampson, Cornish \& McWilliams 2015~\cite{2015PhRvD..91h4055S}).    More recently Kelley et al. (2017)~\cite{2017arXiv170202180K} used the iIllustris simulation to make a new prediction for the GW background signal in the pulsar timing band and their models are not ruled out by the current upper bounds.

SMBHB mergers are not the only possible source of a GW background in the pulsar timing band. Quantum fluctuations of the gravitational field in the early universe, amplified by an inflationary phase could produce a stochastic relic GW background (e.g., Grishchuk 1977~\cite{1977SvPhU..20..319G}). The spectral index of the relic GWs is related to the equation of state of the early universe. By extending the power-law background search to generic spectral indices, Arzoumanian et al. (2016) placed limits on the energy density of relic GWs. From this they obtained limits on the Hubble parameter during inflation. Lasky et al. (2016)~\cite{2016PhRvX...6a1035L} discussed the implications of the bounds on the cosmological background over the entire range of GW frequencies.
 Cosmic strings could also produce a stochastic background of GWs as well as individual bursts (e.g., Damour \& Vilenkin 2001~\cite{2001PhRvD..64f4008D}).  Arzoumanian et al. (2016) placed the most stringent limits to date on a GW background generated by a network of cosmic strings, which translates into a conservative upper limit on cosmic string tension of $G\mu < 3 \times 10^{-8}$ (see also Lentati et al. 2015 and Sanidas, Battye \& Stappers 2012~\cite{2012PhRvD..85l2003S}).   

\section{Current limitations}

Pulsar timing data processing packages are sufficiently advanced to produce pulsar timing residuals at the level needed for GW detection (the \textsc{tempo2} software package already accounts for all known phenomena that affects the propagation of the pulse from the pulsar to the observatory at the 1\,ns level; Hobbs et al. 2006~\cite{2006MNRAS.369..655H}).  For instance, the PTA teams are now achieving sub-100\,ns rms timing residuals on a handful of pulsars (this implies that, for at least a few pulsars, calibration errors, radio interference mitigation, interstellar medium propagation effects, orbital motion and many more phenomena can be accounted for).  GW detection codes have also been well tested on both real and simulated data sets.  The primary issue is that we are simply not achieving the necessary precision for a large sample of pulsars and/or we are not observing enough pulsars.  Jenet et al. (2005)~\cite{2005ApJ...625L.123J} demonstrated that at least 20 pulsars are needed to be observed over 5-10 years with an rms timing residual of 100\,ns in order to make a low-sigma detection of a GW background.  Siemens et al. (2013)~\cite{2013CQGra..30v4015S} considered whether a small number of well timed pulsars or a larger number of poorly-timed pulsars would be preferable and showed that adding in a sufficient number of pulsars with rms residuals of $\sim 1 \mu$s can significantly increase the sensitivity of the array.

In many cases the timing precision is not limited simply by the sensitivity of the observing systems.  For the very brightest pulsars (such as PSR~J0437$-$4715) we are limited by intrinsic pulse variations (a noise process known as ``jitter"; e.g., Shannon et al. 2014~\cite{2014MNRAS.443.1463S}).  Jitter currently only affects a relatively small number of the PTA pulsars, but the era of much larger and more sensitive telescopes jitter will become a major limitation for PTA research. On long time scales many of the pulsars are affected by long-term timing variations known as timing noise (see, e.g., Lentati et al. 2016~\cite{2016MNRAS.458.2161L}, Caballero et al. 2016~\cite{2016MNRAS.457.4421C}, Lam et al. 2017~\cite{2017ApJ...834...35L}) and two of the PTA pulsars (PSRs~J0613$-$0200 and J1824$-$2452A) have been observed to glitch (McKee et al. 2016~\cite{2016MNRAS.461.2809M} and Cognard \& Backer 2005~\cite{2005ASPC..328..389C}, respectively).  Even though mitigation routines exist and wide observing bandwidths are used, the interstellar medium is still a limiting factor for PTAs.  Levin (2015)~\cite{2015JPhCS.610a2020L} reviewed the effects of dispersion and scattering for PTAs.

Ensuring an un-ambiguous detection of a GW signal is also non-trivial. Taylor et al. (2016)~\cite{2016arXiv160609180T} presented two methods for assessing the significance of a stochastic GW background signal in a given data set. Zhu et al. (2015)~\cite{2015MNRAS.449.1650Z} discuss the construction of null streams as a means to provide a consistency check when a single source GW signal is detected.  Tiburzi et al. (2016) showed that, without care, false detections of a GW background could be made even if no GW signal were present.  These false detections can come from incomplete removal (or modelling) of effects such as errors in terrestrial clocks, the solar wind or in the solar system planetary ephemerides. In particular the planetary ephemerides are currently being actively investigated as the more recent ephemerides seem to induce low-frequency noise into the timing residuals (that can mimic the signature of a gravitational wave background; Shannon et al., private communication).  The reason for this is not currently understood, but is being explored by all three PTAs and it is clear that the choice of planetary ephemeris has a significant effect on the resulting data sets and sensitivity to GWs.

\section{What next for pulsar timing arrays?}

The three main PTA projects have now been ongoing for more than a decade.  The data sets  produced are the most high-timing-precision data sets yet made.  Processing such data has required huge improvements in understanding noise processes and calibration methods. The observations have been used to place stringent constraints on the amplitude of GWs. The bounds have been used to constrain cosmological and cosmic string models and to rule out models of GW emission from supermassive binary black holes.    The data have also been used to study atomic clocks and the solar system ephemeris.  Unfortunately GWs have not been detected by the PTAs.  

Determining when we should expect our first detection is non-trivial.  The most recent GW models from supermassive binary black holes predict a background with an amplitude $A \sim 5 \times 10^{-16}$ (e.g., Sesana et al. 2016~\cite{2016MNRAS.463L...6S} and Kelley et al. 2017), but the uncertainty is still large. Predictions from cosmic strings or the early Universe are even more uncertain.  Groups have attempted to calculate the time to detection based on estimates of the GW amplitude (see, e.g., Siemens et al. 2013~\cite{2013CQGra..30v4015S} and Taylor et al. 2016~\cite{2016ApJ...819L...6T}).  The estimated times range from a few years from now to a few decades.  These calculations are made even more uncertain through the assumptions made about how many new pulsars will be discovered (and added into PTAs) over the coming years, what the timing precision will be of those pulsars, whether we will be able to correct for interstellar medium effects at the necessary level, how intrinsically stable the pulsars are and how will new telescopes (that are currently not part of the PTA experiments) contribute.

Some pulsars (such as PSRs~J0437$-$4715, J1713$+$0747 and J1909$-$3744) have significantly smaller timing residuals than many of the other IPTA pulsars.  It is therefore probable that the first evidence of GWs will be in the timing residuals of just a few pulsars. The timing residuals for those pulsars will exhibit low-frequency noise or sinusoidal signatures (caused by the GWs) well before correlations can be measured between a large number of pulsar pairs. Such noise will have two effects: 1) GW bounds will stop decreasing as they will be dominated by the noise in these pulsars and 2) even though pulsar astronomers will not be able to state unambiguously that these residuals have been induced by GWs, statements such as ``if these residuals are induced by GWs then the amplitude and spectral properties of those GWs are \ldots" will become possible.

\subsection{Current telescopes}

Most of the existing telescopes being used for PTA observations will continue to be upgraded.  For instance, the Parkes telescope will soon be commissioning a new ultra-wide-bandwidth receiver covering from 700\,MHz to $\sim$4\,GHz.  This receiver will produce the same observations as the existing receiver suite, but 1) in half the time, 2) with continuous frequency coverage and 3) with cutting-edge radio-frequency-interference mitigation and backend processing.  Similar receiver upgrades are being made at many of the other IPTA telescopes and it is clear that future instrumentation for PTA research will be based around wide-band receivers.  In the near future it is likely that low-frequency observations from e.g., MWA, LOFAR and the GMRT will play an essential role in monitoring the interstellar medium in the direction of the IPTA pulsars (although note the cautions in e.g., Cordes, Shannon \& Stinebring 2016~\cite{2016ApJ...817...16C}).

Many of the IPTA telescopes are also carrying out pulsar surveys in order to detect even more millisecond pulsars to be added into the PTAs. Surveys are currently ongoing with the Parkes (e.g., Keane et al. 2016~\cite{2016arXiv160205165K}), Effelsberg~\cite{2013MNRAS.435.2234B}, Arecibo (e.g., Cordes et al. 2006~\cite{2006ApJ...637..446C}), Green Bank (e.g., Stovall et al. 2014~\cite{2014ApJ...791...67S})  and LOFAR (e.g., Coenen et al. 2014~\cite{2014A&A...570A..60C}) telescopes.  Traditional pulsar search methods are computationally expensive  and this is becoming a limiting factor for surveys on future telescopes.  One recent and very successful method is to observe objects that were identified as gamma-ray emitting sources by the Fermi spacecraft.  To date, such surveys have led to the discovery of more than 75 new millisecond pulsars and many of those are now observed by the PTAs. Various novel search techniques are also being considered which may lead to a much larger sample of IPTA pulsars (see, for instance, Dai et al. 2016, for a description of images of the variance in both time and frequency that can be used to detect pulsars in large-scale continuum surveys)~\cite{2016MNRAS.462.3115D}.

\subsection{New telescopes}

\begin{figure}
\begin{center}
\includegraphics[width=15.5cm]{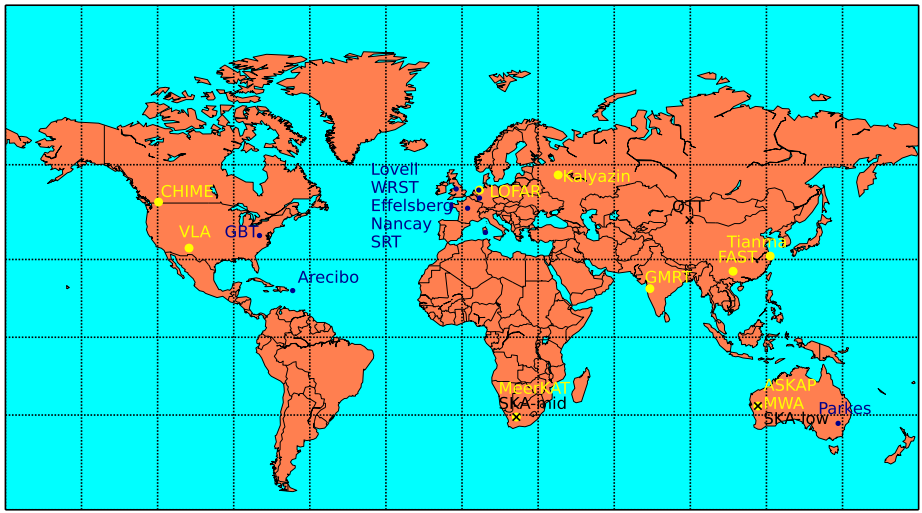}
\caption{Telescopes currently part of the IPTA and those that are expected to contribute soon.  Together these telescopes are able to observe pulsars anywhere in the sky and to detect GWs coming from any direction.  Telescopes labelled in blue are existing IPTA telescopes. Those in yellow exist, but are not officially part of the IPTA yet.  Those labelled in black are being designed.}\label{fg:map}
\end{center}
\end{figure}

A large number of new telescopes will soon to be significantly contributing to PTA projects (see Figure~\ref{fg:map} for a World-map showing current and future PTA telescopes).  The most exciting new developments are the Five Hundred Metre Aperture Spherical Telescope (FAST; e.g., Li et al. 2013~\cite{2013IAUS..291..325L}) that recently opened in China and MeerKAT in South Africa (e.g., Foley et al.. 2016~\cite{2016MNRAS.460.1664F}). Both telescopes will contribute to 1) finding new millisecond pulsars, 2) high-precision pulsar timing and 3) studies of pulsar noise processes that can only be probed by a detailed study of individual pulses (for instance, jitter noise).    

On the longer term the Chinese are also planning a large, single-dish, steerable telescope known as the QiTai (or QTT) telescope (see Xu \& Wang 2016~\cite{2016SPIE.9906E..5LX}). This telescope will operate over a wide range of frequencies and will be able to observe a much larger number of pulsars than is available from FAST.  An analysis of how the Qitai and FAST telescopes will contribute to PTAs was provided in Hobbs et al. (2014)~\cite{2014arXiv1407.0435H}.  In the Southern hemisphere, the Square Kilometre Array will be built in Southern Africa and Australia.  One of the key science goals for the mid-frequency part of this telescope is to detect and study ultra-low-frequency GWs using the methods described in this review (see, e.g., Janssen et al. 2015~\cite{2015aska.confE..37J}).

Over time more and more telescopes will contribute observations to the global IPTA effort.  This means that it will become important to optimise which telescopes observe which pulsars.  Lee et al. (2012)~\cite{2012MNRAS.423.2642L} and Burt, Lommen \& Finn (2011)~\cite{2011ApJ...730...17B} have considered methods for such optimisation.  Of course, such optimisation is non-trivial as pulsar observations are used for so much more than GW detection.  

\section{Conclusion}

Pulsar timing arrays are ongoing with existing telescopes and form key science goals for many of the next generation radio telescopes.  PTAs combine observations of pulsars to probe the population of supermassive black holes, the early universe and cosmic strings.  These are all exciting objects and topics for the general public.  Projects such as PULSE@Parkes (Hobbs et al. 2009)~\cite{2009PASA...26..468H} in which high school students observe pulsars using the Parkes telescope, the Arecibo Remote Command Center (ARCC) project in which students use the Arecibo telescope to search for new pulsars and the Pulsar Search Collaboratory (Rosen et al. 2010)~\cite{2010AEdRv...9a0106R} with Green Bank ensure that the enthusiasm that the IPTA members have for this project gets passed down to the next generation of pulsar astronomers.

Within decades it is likely that ground-based detectors will be analysing high-frequency GWs in detail, space-based detectors will study mid-frequency GWs and PTAs will be probing ultra-low frequency GWs.  Astronomy over the entire spectrum of GW frequencies will soon become a reality.

\section{Acknowledgements}

We thank the following people who read and/or contributed to some part of this paper: Justin Ellis, Paul Lasky, Alberto Sesana, Ryan Shannon, Steve Taylor, Joris Verbiest, Jingbo Wang, Yan Wang and Xingjiang Zhu.    We thank the five reviewers for their useful contributions to the manuscript.

\end{document}